\documentclass[10pt,twoside]{article}
\usepackage{latexsym,amssymb,amsmath,amsthm}
\let\proof\relax
\usepackage{qic}
\usepackage[mathscr]{euscript}
\usepackage[matrix,arrow]{xy}

\begin{document}
\setlength{\textheight}{8.0truein}    %FOR 2ND PAGE ONWARDS

\runninghead{An extremal result for geometries in the one-way measurement model}
            {N.~de~Beaudrap, M.~Pei}

\normalsize\textlineskip
\thispagestyle{empty}
\setcounter{page}{1}

%\copyrightheading{Vol.}{No.}{Year}{Page Nos.}

\vspace*{0.88truein}

\alphfootnote

\fpage{1}

\centerline{\bf	AN EXTREMAL RESULT FOR GEOMETRIES}
\vspace*{0.035truein}
\centerline{\bf	IN THE ONE-WAY MEASUREMENT MODEL}
\vspace*{0.37truein}
\centerline{\footnotesize	NIEL DE BEAUDRAP\footnote{email: jdebeaud@iqc.ca}}
\vspace*{0.015truein}
\centerline{\footnotesize\it Institute for Quantum Computing, University of Waterloo}
\baselineskip=10pt
\centerline{\footnotesize\it  Waterloo, Ontario, N2L 3G1, Canada}
\vspace*{10pt}
\centerline{\footnotesize MARTIN PEI}%\footnote{email: mpei@math.uwaterloo.ca}}
\vspace*{0.015truein}
\centerline{\footnotesize\it Department of Combinatorics and Optimization, University of Waterloo}
\baselineskip=10pt
\centerline{\footnotesize\it Waterloo, Ontario, N2L 3G1, Canada}
\vspace*{0.225truein}
\vspace{4em}
%\publisher{(received date)}{(revised date)}

\vspace*{0.21truein}

\abstracts{%
	We present an extremal result for the class of graphs $G$ which (together with some specified sets
	of input and output vertices, $I$ and $O$) have a certain ``flow'' property introduced by Danos and
	Kashefi for the one-way measurement model of quantum computation. The existence of a flow for
	a triple $(G,I,O)$ allows a unitary embedding to be derived from any choice of measurement bases
	allowed in the one-way measurement model. We prove an upper bound on the number of edges that a
	graph $G$ may have, in order for a triple $(G,I,O)$ to have a flow for some $I, O \subseteq V(G)$,
	in terms of the number of vertices in $G$ and $O$. This implies that finding a flow for a triple
	$(G,I,O)$ when $\lvert I \rvert = \lvert O \rvert = k$ (corresponding to unitary transformations in
	the measurement model) and $\lvert V(G) \rvert = n$ can be performed in time $O(k^2 n)$, improving
	the earlier known bound of $O(km)$ given in \cite{B06a}, where $m = \lvert E(G) \rvert$.
}{}{}

\vspace*{10pt}

\vspace*{3pt}
%\communicate{to be filled by the Editorial}

\vspace*{1pt}\textlineskip    %) USE THIS MEASUREMENT WHEN THERE IS
   %) A SECTION HEADING
%\vspace*{-0.5pt}
%\noindent

% =========================================================================
% ARTICLE-SPECIFIC MACROS.

% Create a new `problem' theorem environment, and override the existing theorem environment.
\let\theorem\relax
\makeatletter
\let\c@theorem\relax
\makeatother
\newtheorem{theorem}{Theorem}
\newtheorem*{problem}{Problem}

\newcommand\N{\mathbb N}
\newcommand\sI{\ensuremath{\mathscr I}}
\newcommand\comp{^{\mathsf c}}

\newcommand\ket[1]{\left|#1\right\rangle}

\newcommand\rulei{\textup{\textrm{\textmd{(i)}}}}
\newcommand\ruleii{\textup{\textrm{\textmd{(ii)}}}}
\newcommand\ruleiii{\textup{\textrm{\textmd{(iii)}}}}
\newcommand\rma{\textup{(\textrm{a})}}
\newcommand\rmb{\textup{(\textrm{b})}}
\newcommand\rmc{\textup{(\textrm{c})}}
\newcommand\rmd{\textup{(\textrm{d})}}
\newcommand\rme{\textup{(\textrm{e})}}
\newcommand\rmf{\textup{(\textrm{f})}}
\newcommand\flowi{(\ref{flowi})}
\newcommand\flowii{(\ref{flowii})}
\newcommand\flowiii{(\ref{flowiii})}
\newcommand\naturali{(\ref{naturali})}
\newcommand\naturalii{(\ref{naturalii})}
\newcommand\naturaliii{(\ref{naturaliii})}

\renewcommand\preceq\preccurlyeq
\renewcommand\ge\geqslant
\renewcommand\le\leqslant
\renewcommand\setminus\smallsetminus

\def\puttext(#1,#2)[#3]#4{\put(#1,#2){\put(-0.5,0){\makebox(0,0)[#3]{#4}}}}

% =========================================================================
% BODY OF THE ARTICLE.

\section{Introduction}

In the one-way measurement model of quantum computation~\cite{RB01,RB02,RBB03,DKP04},
algorithms are described in part by a graph $G$, whose edges $E(G)$ represent entanglement
operations performed on pairs of qubits which are indexed by the set of vertices $V(G)$.
We distinguish two (not necessarily disjoint) sets of vertices, the \emph{input} set $I
\subseteq V(G)$ and the \emph{output} set $O \subseteq V(G)$, which represent subsystems
of qubits which are used to support the input and the output of the algorithm. The qubits of
$I$ may initially be in any state $\ket{\psi}$\,, while the qubits of $I\comp = V(G) \setminus
I$ are initially prepared in the $\ket{+}$ state. We perform controlled-$Z$ operations on pairs
of qubits which are connected by an edge in $G$\,: because these all commute and are symmetric,
order and orientation is unimportant. We then measure each qubit $v$ in $O\comp = V(G) \setminus
O$ in some order, each with a choice of basis $\big\{ \ket{+_{\theta_v}}, \ket{-_{\theta_v}} \big\}$,
where $\ket{\pm_{\theta_v}} \propto \ket{0} \pm \mathrm{e}^{i\theta_v} \ket{1}$.\footnote{These are
antipodal vectors on the Bloch sphere, which lie on the $XY$ plane. The one-way measurement model may
be generalized to allow $XZ$ plane and $YZ$ plane measurements as well; however, we restrict
ourselves here to the original model, in which only $XY$ plane measurements are used.}~ For each
measurement, if we obtain the $\ket{-_{\theta_v}}$ outcome, the angles $\theta_w$ for qubits $w$ which
have not yet measured may undergo a change in sign. Equivalently (albeit less efficiently), we may
perform some correction operation after each measurement which yields a $\ket{-_{\theta_v}}$ outcome
in order to steer the state into what the result would have been had the $\ket{+_{\theta_v}}$ outcome
been produced.

A triple $(G,I,O)$ is called a \emph{geometry}: each measurement-based algorithm has an underlying
geometry, and two distinct algorithms may have the same geometry. A geometry captures the discrete
structure of a one-way measurement algorithm: to develop a theory of discrete structures underlying
quantum algorithms, one may ask whether it is possible to determine which geometries $(G,I,O)$ underlie 
particular classes of operations, e.g. unitary transformations and unitary embeddings. To this end, Danos
and Kashefi~\cite{DK05} introduced the concept of a \emph{flow} as a sufficient condition for a geometry
to underlie a unitary embedding, independent of the measurements to be performed on each qubit of $O\comp$.
(We will refer to these as \emph{causal flows}, in order to distinguish these from e.g. network flows.) 

\vspace{1ex}
\begin{definition}
	\label{dfn:causalFlow}
	Let $(G,I,O)$ be a geometry, and $\sim$ be the adjacency relation in $G$. A \emph{causal flow} on
	a geometry $(G,I,O)$ is an ordered pair $(f, \preceq)$\,, with a function $f: O\comp \to I\comp$
	and a partial order $\preceq$ on $V(G)$\,, such that the relations
	\begin{subequations}
	\begin{align}
			\label{flowi}
			x &	\sim f(x)		
		\\
			\label{flowii}
			x &	\preceq f(x)
		\\
			\label{flowiii}
			y \sim f(x) \;&\implies\; x \preceq y
	\end{align}
	\end{subequations}
	hold for all vertices $x \in O\comp$ and $y \in V(G)$\,.
\end{definition}
\vspace{1ex}

\begin{figure}[h]
	% Fix QIC's figure caption command for this figure
	% so that side-by-side captions don't overwrite each other
	\renewcommand{\fcaption}[1]{
        \refstepcounter{figure}
             {\begin{center}
					\vspace{2ex}
					\begin{minipage}{\textwidth}
	             {\footnotesize Fig.~\thefigure. #1}
					\end{minipage}
              \end{center}}}
% 
% 	Figure 1
	\begin{minipage}[c]{0.55\textwidth}
		\setlength\unitlength{0.133mm}
		\framebox{
		\hspace{0.05\textwidth}
		\begin{picture}(240,300)(10,0)
			\put(-5,140){
			{\includegraphics[width=20mm]{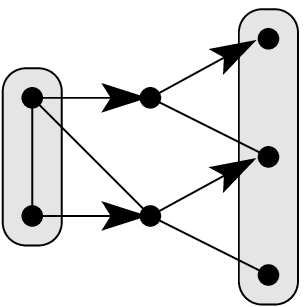}}
			\puttext(-140,15)[c]{$I$}
			\puttext(-13,-15)[c]{$O$}
			\puttext(-165,105)[c]{$a$}
			\puttext(-165,50)[c]{$b$}
 			\puttext(-75,125)[c]{$c$}
 			\puttext(-75,25)[c]{$d$}
 			\puttext(12,138)[c]{$g$}
 			\puttext(12,78)[c]{$h$}
 			\puttext(12,18)[c]{$\ell$}
			\put(-205,-55){
				\xymatrix @R=0ex @!C=0.5ex{
							  													& a \ar@{-}[r]\ar@{-}[rd]	& c \ar@{-}[r]	&	g	\\
								b \ar@{-}[ur]\ar@{-}[r]\ar@{-}[rd]	& d \ar@{-}[ur]\ar@{-}[r]	&	h	\\
																				&	\ell
			}}}
		\end{picture}
 		\hspace{-0.12\textwidth}
		\vline
 		\hspace{0.03\textwidth}
		\begin{picture}(200,100)(-5,0)
			\put(-7,180){
			{\includegraphics[width=16mm, angle=-45]{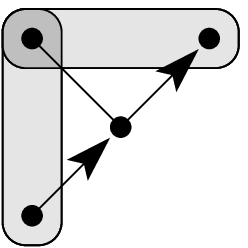}}
			\puttext(-150,52)[c]{$I$}
			\puttext(-25,52)[c]{$O$}
			\puttext(-178,0)[c]{$a$}
			\puttext(-85,95)[c]{$b$}
			\puttext(-85,-16)[c]{$c$}
			\puttext(6,1)[c]{$d$}
			}
			\put(-8,75){
				\xymatrix @R=0ex @!C=0.5ex{
							  									& b 						\\
								a \ar@{-}[ur]\ar@{-}[r]	& c \ar@{-}[r]	&	d
			}}
		\end{picture}
		\vline\/
		\hspace{0.04\textwidth}
     	\begin{picture}(55,140)
		\put(-5,160){
			{\includegraphics[width=4mm]{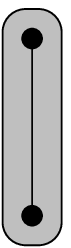}}
			\puttext(-17,-18)[c]{$I = O$}
			\puttext(-43,97)[c]{$a$}
			\puttext(-43,19)[c]{$b$}}
		\put(-10,80){
				\xymatrix @R=0ex @!C=0.5ex{ a \\ \\ b }}
		\end{picture}
		}%
		\fcaption{\label{fig:examplesFlow}
				Examples of geometries with causal flows. The arrows in each indicates the action of
				a function $f: O\comp \longrightarrow I\comp$\,,\/ along \emph{undirected} edges.
 				Compatible partial orders $\preceq$ for each example are given by Hasse diagrams
 				(read from left to right).}
	\end{minipage}
	\hfill
	\begin{minipage}[c]{0.39\textwidth}\setlength\unitlength{0.133mm}
		\framebox{
 			\begin{picture}(390,300)(0,0)
			\put(100,140){
				\includegraphics[width=20mm]{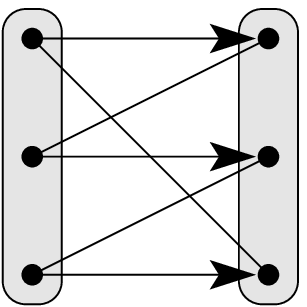}
 				\puttext(-140,-20)[c]{$I$}
 				\puttext(-20,-20)[c]{$O$}
 				\puttext(-165,140)[r]{$a_0$}
 				\puttext(-165,80)[r]{$a_1$}
 				\puttext(-165,20)[r]{$a_2$}
 				\puttext(30,140)[r]{$b_0$}
 				\puttext(30,80)[r]{$b_1$}
 				\puttext(30,20)[r]{$b_2$}
			}
 			\put(-20,70){
 				\xymatrix @R=0.1ex @!C=0.5ex{
 					&										&	b_0								&	b_1								&	b_2								& b_0\\
 					\cdots\ar@{-}[r] &	{a_0}\ar@{-}[ur]\ar@{-}[r]	&	a_1\ar@{-}[ur]\ar@{-}[r]	&	a_2\ar@{-}[ur]\ar@{-}[r]	&	a_0\ar@{-}[ur]\ar@{-}[r]	&	\cdots\\
 					&										&										&										&
 				}}
 			\end{picture}}
			\vspace{-0.85em}
			\fcaption{\label{fig:exampleNoFlow}
					A geometry with no causal flows. Illustrated here is a particular 
					injection $f: O\comp \longrightarrow I\comp$, and the coarsest pre-order satisfying
					\flowii\ and \flowiii\/.}
	\end{minipage}
\end{figure}

Figures~\ref{fig:examplesFlow} and~\ref{fig:exampleNoFlow} illustrate examples of geometries with and
without causal flows.

For any choice of measurement operations on the qubits of $O\comp$, a causal flow for a geometry determines
an order in which the measurements may be performed, and corresponding correction operations, so that the
resulting algorithm performs a unitary embedding from (the initial state of) the input subsystem $I$ to
(the final state of) the output subsystem $O$~\cite{DK05}. A causal flow then allows a partial specification
of a unitary transformation in the one-way measurement model to be interpolated into a complete algorithm.
The ability to find flows for arbitrary geometries suggests that new techniques may be developed
(such as suggested in~\cite{BDK06} and~\cite{BK07}) to devise and analyze quantum algorithms in the
one-way measurement model.

For the special case where $\lvert I \rvert = \lvert O \rvert$, corresponding e.g. to measurement
patterns which implement unitary transformations, it is possible to efficiently determine whether
a geometry $(G,I,O)$ has a casual flow, and to construct one in the case that a flow exists~\cite{B06a}.
This problem can be reduced to the Maximum Flow and Transitive Closure problems on digraphs: using
standard algorithms to solve these problems, finding a causal flow for $(G,I,O)$ when $\lvert I \rvert =
\lvert O \rvert = k$ and $\lvert E(G) \rvert = m$ can be solved in time $O(km)$. 
 
In this paper, we present an extremal result: in a geometry $(G,I,O)$ which has a causal flow, with
$\lvert V(G) \rvert = n$ and $\lvert O \rvert = k$, the maximum number of edges that $G$ may have
is $kn - \binom{k+1}{2}$. This allows the running time of the algorithm of \cite{B06a} to be
improved to $O(k^2 n)$, by rejecting graphs with more than $kn - \binom{k+1}{2}$ edges as a
preliminary step. It also implies that no improvement can be made in terms of a stronger upper
bound on the number of edges in the geometry as a whole.

\subsection*{Notation and conventions.}

We will represent an (undirected) edge between vertices $x$ and $y$ in a graph by $xy$\,, and directed edges 
(or \emph{arcs}) in a directed graph (or \emph{digraph}) by $x \to y$. Directed paths and cycles will be
represented by sequences of arcs, $x \to y \to z \to \cdots\;$. We use the convention that digraphs may contain
loops on a single vertex and multiple edges between two vertices, but that graphs cannot have either.

%=================================================================================
\section{Characterization of causal flows in terms of paths and digraphs}
\label{sec:digraphFlowConditions}

In this section, we outline the characterization of causal flows in terms of
collections of paths and acyclic digraphs described in~\cite{B06a}. This will allow us
to abstract away some details of geometries which are not essential to the analysis, and
simplify the proof of the extremal result.

\subsection{Vertex-disjoint paths in place of flow-functions}

In a causal flow $(f,\preceq)$ on a geometry $(G,I,O)$, it is easy to show that the function $f$
must be injective: for two vertices $x, y \in O\comp$, if $f(x) = f(y)$, then $x \sim f(y)$
by the relation \flowi, and thus $x \preceq y$ by the relation \flowiii\,; and similarly $y \preceq x$,
in which case $x = y$. Then, the orbits of the function $f$ are a collection $C$ of (possibly trivial)
directed paths, which are vertex disjoint by the injectivity of $f$. More precisely, we have:

\vspace{1ex}
\begin{lemma}[{\cite[Lemma 3]{B06a}}]
	\label{lemma:pathCover}
	Let $(f, \preceq)$ be a causal flow on a geometry $(G,I,O)$\,. Then there is a collection of (possibly
	trivial) directed paths $P_1, \ldots, P_k$ in $G$ such that the following hold:
	\begin{romanlist}
	\item
		each $v \in V(G)$ is contained in exactly one path $P_j$;
	\item
		each path $P_j$ is either disjoint from $I$\,, or intersects $I$ only at its
		initial point;
	\item
		each path $P_j$ intersects $O$ only at its final point;
	\item
		there is an arc $x \to y$ in some path $P_j$ iff $y = f(x)$.
	\end{romanlist}
\end{lemma}
\vspace{1ex}

That each path ends at a vertex in $O$ is easy to show: because any vertex of $O\comp$ is in the domain of
$f$, the final points of each dipaths can only be a vertex of $O$\,. Also, every vertex of $O$ is in such a
path, if only a trivial path (i.e. one of length zero, consisting of just that vertex). Thus, there are as
many paths in the collection as there are vertices of $O$. Note that in general, there may be more paths than
there are input vertices. All of the above observations are illustrated in the examples of Figure~\ref{fig:examplesFlow},
taking the arrows to represent the edges of the directed paths induced by $f$.

For the question of whether a geometry on $n$ vertices and with $m$ edges has a causal flow, we may consider
any graph $G$ with these properties, under the constraint that it admits some family of vertex-disjoint (directed)
paths $P_1,\ldots,P_k$ covering the entire graph. Without loss of generality, we may then let $O$ be the final points
of those paths, and $I$ be an arbitrary subset of the initial points of those paths. We may then consider whether the
function $f$ mapping each (non-terminal) vertex to its' successor in its' respective path is consistent with some
partial order $\preceq$ to form a causal flow $(f,\preceq)$\,.

\subsection{An acyclic digraph construction for the partial order}

Given a candidate function $f$, there is a natural choice of binary relation to consider in order to determine
whether $f$ is part of a causal flow:

\vspace{1ex}
\begin{definition}
	\label{dfn:naturalPreorder}
	Let $(G,I,O)$ be a geometry, and $f: O\comp \longrightarrow I\comp$ be an injective function. The \emph{natural 
	pre-order\footnote{A pre-order is a binary relation which is reflexive and transitive, but not necessarily
	antisymmetric.}~ $\preceq$ for $f$} is the transitive closure on $V(G)$ of the conditions
	\begin{subequations}
	\begin{align}
			\label{naturali}
			x &\preceq x
		\\
			\label{naturalii}
			x &\preceq f(x)
		\\
			\label{naturaliii}
			y \sim f(x) \;\;&\implies\;\; x \preceq y
	\end{align}
	\end{subequations}
	for all $x, y \in V(G)$\,.
\end{definition}
\vspace{1ex}

By definition, the natural pre-order $\preceq$ for a function $f$ is the coarsest reflexive and transitive
binary relation which satisfies the causal flow relations \flowii\ and \flowiii: thus, if $f$ is a part of
any causal flow, then $(f,\preceq)$ in particular will be a causal flow.

Given a collection of vertex-disjoint dipaths $P_1,\ldots,P_k$ characterizing a function $f$, rather than
consider the natural pre-order $\preceq$ of $f$ explicitly, we may characterize $\preceq$ in terms of a digraph
containing the paths $P_1,\ldots,P_k$\,. Following~\cite{B06a}, given an injective function $f$ from such a family
of paths, we may construct the \emph{influencing digraph} $\sI_f$ on the vertices $V(G)$\,, where $x \to y$ is
an arc of $\sI_f$ if one of $y = x$, $y = f(x)$\,, or $y \sim f(x)$ holds in $G$. (Equivalently: we have $x \to y$
an arc of $\sI_f$ for distinct $x$ and $y$ either if $y$ is the successor of $x$ in a path $P_i$, or if
$y$ is adjacent in $G$ to the successor of $x$ in a path $P_i$.) The latter two conditions correspond to the
relations \naturalii\ and \naturaliii: then, for the natural pre-order $\preceq$ for $f$, we have $x \preceq y$
whenever there is a directed path from $x$ to $y$ in $\sI_f$\,.

In order for $\preceq$ to be a partial order, it must be anti-symmetric, which is equivalent to $\sI_f$
having no circuits (except for circuits which repeatedly visit the same vertex by traversing the loop at
that vertex). We may simplify this by considering the digraph obtained by deleting the loops from $\sI_f$\,.
Expressed in terms of families of dipaths $P_1,\ldots,P_k$, this digraph is:

\vspace{1ex}
\begin{definition}
	\label{dfn:assocDigraph}
	Let $G$ be a graph with vertex-disjoint dipaths $P_1,\ldots,P_k$ covering $V(G)$.
	Then $D(G,P_1,\ldots,P_k)$ is the digraph such that $x \to y$ is an arc of $D$ if and only
	if either there is a path $P_j$ which contains an arc $x \to y$, or or there is a vertex
	$z \in V(G)$ and a path $P_j$ such that $x \to z$ is an arc of $P_j$ and $y \sim z$ in $G$.
\end{definition}
\vspace{1ex}

Then, the injective function $f$ induced by the family of dipaths $P_1,\ldots,P_k$ is consistent
with a causal flow (again taking the inputs and outputs to the be endpoints of the paths $P_j$)
if and only if the digraph $D(G,P_1,\ldots,P_k)$ is acyclic.

\section{Analysis of the extremal problem}

The extremal problem to be solved is: given that a geometry $(G,I,O)$ has a causal flow, what is
the maximum number of edges that $G$ may have, given that $\lvert V(G) \rvert = n$ and $\lvert O \rvert = k$,
in terms of $n$ and $k$? Using the characterizations of the previous section of causal flows, and the
graphs of geometries which have causal flows, we may rephrase this problem as follows:

\vspace{1ex}
\begin{problem}
	Let $n,k$ be integers where $n \ge k$.  Let $G$ be a graph on $n$ vertices which includes $k$ vertex-disjoint
	(directed) paths $P_1,\ldots,P_k$ that cover $V(G)$, and let $D(G,P_1,\ldots,P_k)$ be the digraph given in 
	Definition~\ref{dfn:assocDigraph}. What is the maximum number of edges $\Gamma(n,k)$ that $G$ may have, under
	the constraint that $D(G,P_1,\ldots,P_k)$ is acyclic?
\end{problem}

\vspace{1ex}

\begin{theorem}
	\label{thm:extremalResult}
	$\Gamma(n,k) = kn - \binom{k+1}{2}$ for all integers $n \ge k \ge 1$.
\end{theorem}
\vspace{1ex}

In this section, we prove this Theorem by bounding the number of
edges between any two paths $P_i$ and $P_j$, and then provide a construction which saturates this bound.

\subsection{Upper bound}

To provide an upper bound on $\Gamma(n,k)$, we make the following observations. Let $G$ and
$P_1,\ldots,P_k$ be as described in the problem above, and let $D = D(G,P_1,\ldots,P_k)$.

\vspace{-1.5ex}
\paragraph{Observation 1.} Consider any one of the paths $P_i = v_1 \to v_2 \to \cdots \to v_{n_i}$. If
$D$ is acyclic, then $v_a v_b \in E(G)$ for $a < b$ if and only if $b = a+1$. (Otherwise, $D$ would
contain the directed cycle $v_a \to v_{a+1} \to \cdots \to v_{b-1} \to v_a$\,, contrary to hypothesis.)
\vspace{-1.5ex}

\paragraph{Observation 2.} Consider any two distinct paths $P_i = v_1 \to v_2 \to \cdots \to v_{n_i}$ and
$P_j = w_1 \to w_2 \to \cdots \to w_{n_j}$\,. If $D$ is acyclic, then there cannot be two edges
$v_a w_b\,,\, v_c w_d \in E(G)$ where $a<c$ and $b>d$. (Otherwise, $D$ would contain the directed cycle
$v_a \to \cdots \to v_{c-1} \to w_d \to \cdots \to w_{b-1} \to v_a$\,, contrary to hypothesis.)
\vspace{1ex}

The first observation indicates that aside from the edges contained in the paths $P_i$ themselves, the
only edges $G$ may have are between pairs of paths, which we will \emph{connecting edges}. The second
observation imposes a constraint on the connecting edges that may exist between any two paths. We may use
these to obtain:

\vspace{1ex}
\begin{lemma}
	\label{lemma:upperBound}
	$\Gamma(n,k) \le kn - \binom{k+1}{2}$ for all integers $n \ge k \ge 1$.
\end{lemma}
\vspace{1ex}

\proof{Consider a graph $G$ and dipaths $P_1, \ldots, P_k$ as above, where each path $P_i$ has $n_i$ vertices,
such that $D(G,P_1,\ldots,P_k)$ is acyclic. We will proceed by bounding the number of connecting edges in $G$
that may exist between each pair of paths $P_i$ and $P_j$\,.

Define a function $\lambda$ from the connecting edges of $G$ to the integers as follows. For any connecting
edge $v_a w_b$, where $v_a$ is the $a^{\textrm{th}}$ vertex of some path $P_i$ and $w_b$ be the $b^{\textrm{th}}$
vertex of some path $P_j$, let $\lambda(v_a w_b) = a+b$\,. Consider two distinct connecting edges $v_a w_b\,,\,
v_c w_d \in E(G)$ between the same two paths $P_i$ and $P_j$\,. By Observation 2, if $a < c$, then $b \le d$\,,
and if $a > c$, then $b \ge d$. In any case, we have $\lambda(v_a w_b) = a+b \ne c+d = \lambda(v_c w_d)$: that
is, each connecting edge between $P_i$ and $P_j$ has a different image in the function $\lambda$. Because
$2 \le \lambda(e) \le n_i + n_j$ for a connecting edge $e$ between $P_i$ and $P_j$, there are at most
$n_i + n_j - 1$ connecting edges between $P_i$ and $P_j$.

Applying this to all pairs of paths $P_i$ and $P_j$, the number of connecting edges in $G$ is then
bounded above by
\begin{align}
		\sum_{1\le i < j \le k} (n_i+n_j-1)
	\;\;=&\;\;
		\tfrac{1}{2}\left[ \sum_{i = 1}^k \sum_{j = 1}^k (n_i + n_j - 1) \;\;-\;\; \sum_{i = 1}^k (2n_i - 1) \right]
	\notag\\=&\;\;
		\tfrac{1}{2}\left[ \sum_{i = 1}^k (k n_i + n - k) \;\;-\;\; \sum_{i = 1}^k (2n_i - 1) \right]
	\notag\\=&\;\;
 		kn - n - \tfrac{1}{2}(k^2 - k)	\;.
\end{align}
As the number of edges in the paths $P_i$ themselves is $n - k$, the total number of edges $G$ may have
is at most $kn - k - \tfrac{1}{2}(k^2 - k) \;=\; kn - \binom{k+1}{2}$}

\subsection{Lower bound}
\label{sec:lowerBound}

\begin{figure}[h]
	\begin{center}
		\includegraphics[width=.8\textwidth]{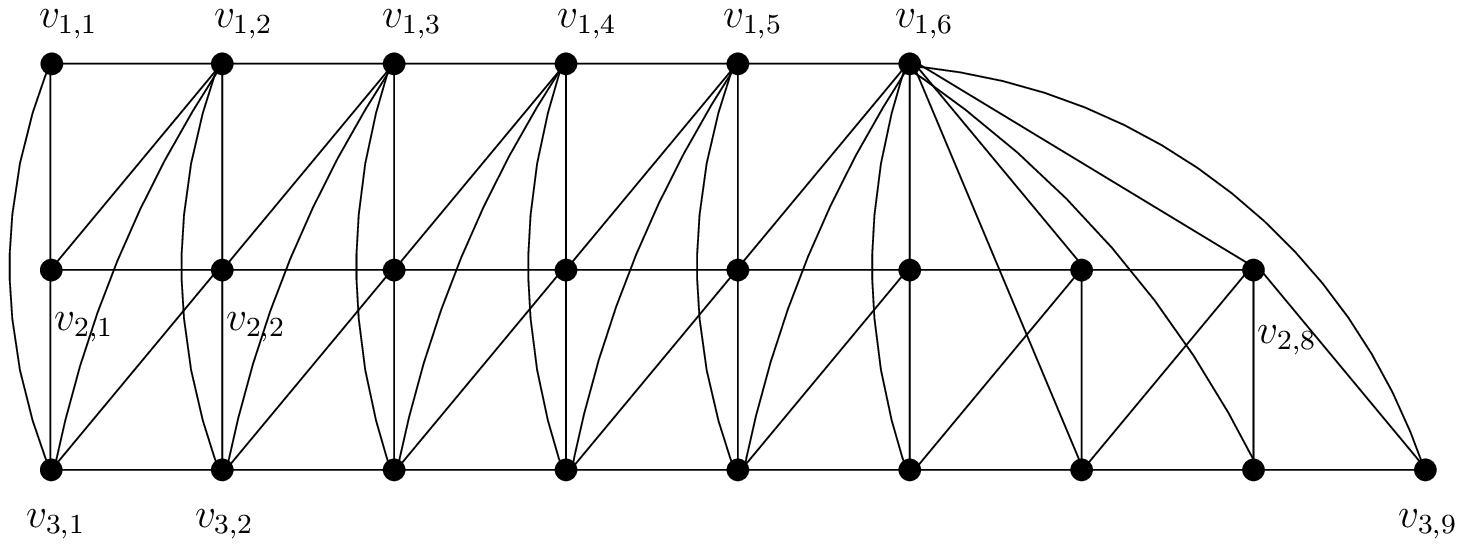}
	\end{center}
	\fcaption{The graph $G(n_1,n_2,n_3)$ for $n_1=6$, $n_2=8$, $n_3=9$.}
	\label{fig:example}
\end{figure}

Consider the following construction for any $n$ and $k$. Let $n_1,n_2,\ldots,n_k$ be an integer partition
of $n$ such that $n_1 \le n_2 \le \cdots \le n_k$\,. For each $1 \le i \le k$\,, let $P_i \,=\, v_{i,1}\,
v_{i,2}\,\cdots\,v_{i,n_i}$\,. Define $G(n_1,\ldots,n_k)$ to be the graph containing these paths, as well
as the following connecting edges for each $1 \le i < j \le k$\,:
\begin{romanlist}
\item
	If $n_i > 1$, then for each $1 \le a < n_i$\,, we include the edge $v_{i,a} v_{j,a}$;
	
\item
	If $n_j > 1$, then for each $1 \le a < n_i$\,, we include the edge $v_{i,a+1} v_{j,a}$;

\item
	For each $n_i \le a \le n_j$\,, we include the edge $v_{i,n_i} v_{j,a}$.
\end{romanlist}
An example of this construction for $k=3$ and an integer partition of $n = 23$ is illustrated in Figure~\ref{fig:example}.

We will describe \rulei~--~\ruleiii\ as the \emph{types} of connecting edges in $G = G(n_1,\ldots,n_k)$\,. Each
connecting edge in $G$ induces up to two arcs in the associated digraph $D = D(G,P_1,\ldots,P_k)$\,. For each
$1 \le i < j \le k$\,, the arcs of $D$ induced by connecting edges between the paths $P_i$ and $P_j$ are of
six different types, labelled here from \rma\ to \rmf, which we group together by the type of the connecting edge which induce them:
\begin{romanlist}
\item	$
	\begin{cases}
		\;\rma\;\;	v_{i,a-1} \to v_{j,a}	&\;\;\;\:\text{for $1 < a \le n_i$ (if $n_i > 1$), and}	\\
		\;\rmb\;\;	v_{j,a-1} \to v_{i,a}	&\;\;\;\:\text{for $1 < a \le n_i$ (if $n_j > 1$);}
	\end{cases}$
	\vspace{1ex}

\item	$
	\begin{cases}
		\;\rmc\;\;	v_{i,a} \to v_{j,a}		&\text{for $1 \le a < n_i - 1$ (if $n_i > 1$), and}	\\
		\;\rmd\;\;	v_{j,a-1} \to v_{i,a+1}	&\text{for $1 < a \le n_i - 1$ (if $n_i > 1$);}
	\end{cases}$
	\vspace{1ex}
	
\item	$
	\begin{cases}
		\;\rme\;\;	v_{i,n_i-1} \to v_{j,a}		&\;\:\,\text{for $n_i \le a \le n_j$, and}	\\
		\;\rmf\;\;	v_{j,a-1} \to v_{i,n_i}		&\;\:\,\text{for $\max \{ n_i \,, 2 \} \le a \le n_j$ (if $n_j > 1$).}
	\end{cases}$
\end{romanlist}
In addition to these arcs induced by connecting edges, $D$ also contains arcs $v_{i,a} \to v_{i,a+1}$ from orienting
the paths $P_i$ themselves.

\vspace{1ex}
\begin{lemma}
	\label{lemma:acyclicConstr}
	The digraph $D = D(G,P_1,\ldots,P_k)$ described above is acyclic.
\end{lemma}

\vspace{1ex}
\proof{
	The arcs in $D$ produced by the rules \rma~--~\rme\ are either of the form $v_{i,a} \to v_{j,b}$ with $a < b$ and no
	constraints on $i$ and $j$\,, or $v_{i,a} \to v_{j,a}$ with $i < j$\,. In either case, if an arc $v_{i,a} \to v_{j,b}$
	is of one of the types \rma~--~\rme, we have $(a,i) < (b,j)$ in the lexicographic ordering on ordered pairs of integers.
	The same also holds for the arcs $v_{i,a} \to v_{i,a+1}$ of the paths $P_i$. Then, if there are arcs in $D(G,P_1,\ldots,P_k)$
	for $v_{j,b} \to v_{i,a}$ where $(b,j) > (a,i)$\,, they must arise from the rule~\rmf, in which case $a = n_i$\,.

	Note that none of the rules \rma~--~\rmf\ produce arcs which leave the final vertex $v_{i,n_i}$ of any path $P_i$\,,
	so there are no non-trivial walks in $D$ which leave such a vertex. Then, it is easy then to show by induction that if
	there is a directed walk in $D$ between distinct vertices $v_{i,a}$ and $v_{j,b}$\,, either $(a,i) < (b,j)$ in the
	lexicographic order, or $b = n_j$\,.

	Let $v_{i,a}$ and $v_{j,b}$ be two vertices, with a directed walk $W$ from $v_{i,a}$ to $v_{j,b}$\,. Because of
	the existence of $W$, we know that $a \ne n_i$\,; then, there is a directed walk from $v_{j,b}$ to $v_{i,a}$ only
	if $(b,j) < (a,i)$\,. We would then have $b = n_j$\,, in which case there are no directed walks from $v_{j,b}$
	to \emph{any} other vertices in $D$\,. So, for any two distinct vertices $v_{i,a}$ and $v_{j,b}$\,, there cannot
	be a directed walk from $v_{i,a}$ to $v_{j,b}$ and also from $v_{j,b}$ to $v_{i,a}$\,, in which case $D$ is acyclic}
\vspace{1ex}

As well as giving rise to an acyclic digraph $D(G,P_1,\ldots,P_k)$\,, we also have:

\vspace{1ex}
\begin{lemma}
	\label{lemma:constrMaximal}
	$\big\lvert E\big(G(n_1,\ldots,n_k)\big) \big\rvert \,=\, kn - \binom{k+1}{2}$\,,\, for any $n \ge k \ge 1$ and integer
	partition $n_1 \le \cdots \le n_k$ of $n$.
\end{lemma}
\vspace{1ex}
\proof{
	Between any pair of paths $P_i$ and $P_j$ in $G(n_1,\ldots,n_k)$\,, there are $n_i - 1$ connecting edges of type
	\rulei, $n_i - 1$ connecting edges of type \ruleii, and $n_j - n_i + 1$ connecting edges of type \ruleiii. There are
	then $n_i + n_j - 1$ connecting edges between $P_i$ and $P_j$\,. This saturates the upper bound for connecting
	edges between pairs of paths in Lemma~\ref{lemma:upperBound}: summed over all pairs of paths, and including
	the edges in the paths $P_i$, the total number of edges in $G(n_1,\ldots,n_k)$ is then $kn - \binom{k+1}{2}$}
\vspace{1ex}

Theorem~\ref{thm:extremalResult} then follows from Lemmas~\ref{lemma:upperBound} and~\ref{lemma:constrMaximal}.
Together with the characterizations described in Section~\ref{sec:digraphFlowConditions}, we then have the result:

\vspace{1ex}
\begin{theorem}
	If a geometry $(G,I,O)$ has a causal flow, then $\lvert E(G) \rvert \,\le\, kn \;-\; \binom{k+1}{2}$\,, where
	$n = \lvert V(G) \rvert$ and $k = \lvert O \rvert$; and the geometry $(G,I,O)$ given by $G = G(n_1,\ldots,n_k)$\,,
	with $I = \big\{v_{i,1}\big\}_{i = 1}^k$ and $O = \big\{v_{i,n_i}\big\}_{i = 1}^k$\,,
	saturates this bound for any partition $n_1 \le \cdots \le n_k$ of $n$.
\end{theorem}

\section{Remarks and Open Problems}

This paper addresses an open problem of~\cite{B06b}, which asked whether a construction similar
to that of Section~\ref{sec:lowerBound} had the maximum possible number of edges for a geometry
having a causal flow, on $n$ vertices and $k$ output vertices.

This extremal result allows us to derive an improved upper bound on the time complexity given by~\cite{B06a}
for recognizing geometries $(G,I,O)$ with flows for the special case $|I| = |O|$\,: by adding a
preliminary step where $|E(G)|$ is compared to $\Gamma(n,k)$, we can quickly eliminate geometries
with too many edges, and perform the rest of the algorithm of~\cite{B06b} for geometries with
$|E(G)| \le \Gamma(n,k)$\,. This yields a running time of $O(k^2 n)$ for finding causal flows
(or determining that none exist) in the case $\lvert I \rvert = \lvert O \rvert = k$. As well,
although there is no known efficient algorithm for determining whether a geometry $(G,I,O)$
has a causal flow in the case $\lvert O \rvert > \lvert I \rvert$, comparing $\lvert E(G) \rvert$
to $\Gamma(n,k)$ provides a simple check which can show that some geometries cannot have a flow
without having to analyze their structure.

Can these techniques can be generalized beyond the special case of causal flows?
The presence of a causal flow is a sufficient \emph{but not a necessary} condition for a geometry $(G,I,O)$
to underlie a unitary embedding in the one-way measurement model~\cite{DK05,BKMP07}: a more general class
of geometries which underlie unitary embeddings are ones with ``generalized flows'', as defined in
\cite{BKMP07}. Obtaining bounds on the number of edges in a geometry with a generalized flow (with or
without imposing constraints such as, e.g. that all measurement operators are $XY$ plane measurements)
would be a step towards more general algorithms for determining when geometries underlie unitary operations
in the one-way measurement model.

\nonumsection{Acknowledgements}
\noindent
We would like to thank Penny Haxell for arranging our collaboration.
N~de~B would also like to thank Elham Kashefi for posing him the problem of ways to determine that a geometry
has no causal flow, without having complete information about the graph.

% =========================================================================
\nonumsection{References}

% =========================================================================
\end{document}